\documentclass[11pt,a4paper]{article}
\usepackage{graphicx}
\usepackage{amsfonts}
\usepackage{amssymb}
\usepackage{amscd}

\def\beq{\begin{equation}}
\def\eeq{\end{equation}}
\def\bea{\begin{eqnarray}}
\def\eea{\end{eqnarray}}
\def\ba{\begin{array}}
\def\ea{\end{array}}
\def\bi{\begin{itemize}}
\def\ei{\end{itemize}}

\begin{document}

\title{LAGRANGIANS WITH \\ RIEMANN ZETA FUNCTION}

\author{ BRANKO DRAGOVICH\\
Institute of Physics
\\Pregrevica 118, P.O. Box 57, 11001 Belgrade, Serbia\\
E-mail: dragovich@phy.bg.ac.yu}

\date{}
\maketitle
\begin{center}
(Received \today )
\end{center}

\begin{abstract}
We consider construction of some Lagrangians which contain the
Riemann zeta function. The starting point in their construction is
$p$-adic string theory. These Lagrangians describe some nonlocal and
nonpolynomial scalar field models, where nonlocality is controlled
by the operator valued Riemann zeta function. The main motivation
for this research is intention to find an effective Lagrangian for
adelic scalar strings.
\end{abstract}

{\centering\section{INTRODUCTION \label{int} }}

 Since 1987, $p$-adicity  has been found in string theory as well as
 in many other models of modern mathematical physics (for an early review, see e.g. \cite{freund1,volovich2}).
One of the most important results in $p$-adic string theory is
construction of an effective Lagrangian for open scalar $p$-adic
string tachyon \cite{freund2,frampton1}. This simple Lagrangian
describes four-point scattering amplitudes as well as all higher
ones at the tree-level.

Using this Lagrangian many aspects of $p$-adic string dynamics have
been investigated and compared with dynamics of ordinary strings.
Mathematical study of spatially homogeneous solutions of the
relevant nonlinear differential equation of motion has been of
particular interest (see \cite{zwiebach,vladimirov2} and references
therein). Some possible cosmological implications of $p$-adic string
theory have been also investigated \cite{barnaby2,arefeva4}. As a
result of these developments,  there have been established many
similarities and analogies between $p$-adic and ordinary strings.

Adelic approach to the string scattering amplitude connects its
$p$-adic and ordinary counterparts (see \cite{freund1, volovich2} as
a review). It gives rise to introduce adelic strings, which contain
ordinary and $p$-adic strings as particular adelic states. Adelic
generalization of quantum mechanics was also successfully
formulated, and it was found a connection between adelic vacuum
state of the harmonic oscillator and the Riemann zeta function
\cite{dragovich2}. Roughly speaking, adelic description is more
profound than real or p-adic ones separately. Consequently, adelic
strings are more fundamental objects than ordinary ones, described
by real numbers.

The present paper can be regarded as a result of some attempts to
construct an effective Lagrangian for adelic scalar string. Our
approach is as follows: we start with the exact Lagrangian for the
effective field of $p$-adic tachyon string, extend prime number $p$
to arbitrary natural number $n$, undertake various summations of
such Lagrangians over  $n$, and obtain some scalar field models with
the operator valued Riemann zeta function. This zeta function
controls spacetime nonlocality. These scalar field models are also
interesting in their own write.

\bigskip

{\centering\section{CONSTRUCTION OF ZETA LAGRANGIANS \label{cr}}}

The exact tree-level Lagrangian of effective scalar field $\varphi$
for open $p$-adic string tachyon is

\begin{equation} {\cal L}_p = \frac{m_p^D}{g_p^2}\, \frac{p^2}{p-1} \Big[
-\frac{1}{2}\, \varphi \, p^{-\frac{\Box}{2 m_p^2}} \, \varphi  +
\frac{1}{p+1}\, \varphi^{p+1} \Big]\,,  \label{2.1} \end{equation}
where $p$
 is any prime number, $\Box = - \partial_t^2  + \nabla^2$ is the
$D$-dimensional d'Alembertian.

The equation of motion for (\ref{2.1}) is

\begin{equation} p^{-\frac{\Box}{2 m_p^2}}\, \varphi = \varphi^p \,,
\label{2.2} \end{equation} and its properties have been studied by
many authors (see, e.g. \cite{zwiebach,vladimirov2} and references
therein).

Prime number $p$ in (\ref{2.1}) and (\ref{2.2}) can be replaced by
any natural number $n \geq 2$ and such expressions also make sense.
Moreover,  taking $p = 1 + \varepsilon \to 1$  there is the limit of
(\ref{2.1})

\begin{equation} {\cal L}_0 = \frac{m^D}{g^2}\,  \Big[ \frac{1}{2}\, \varphi \,
\frac{\Box}{m^2} \, \varphi  + \frac{\varphi^2}{2}\, ( \ln
\varphi^{2} -1 ) \Big]\, \label{2.3}\end{equation}
 which corresponds to the ordinary bosonic string in the boundary string field
theory \cite{gerasimov}.

Now we want to introduce a Lagrangian which incorporates all the
above  Lagrangians (\ref{2.1}), with $p$ replaced by $n \in
\mathbb{N}$, and ${\cal L}_0$ (\ref{2.3}). To this end, we take the
sum of all Lagrangians ${\cal L}_n$  in the form

\bea L =   \sum_{n = 0}^{+\infty} C_n\, {\cal L}_n   = C_0\, {\cal
L}_0 + \sum_{n= 1}^{+\infty} C_n \frac{ m_n^D}{g_n^2}\frac{n^2}{n
-1} \Big[ -\frac{1}{2}\, \phi \, n^{-\frac{\Box}{2 m_n^2}} \, \phi +
\frac{1}{n + 1} \, \phi^{n+1} \Big]\,, \label{2.4} \eea whose
explicit realization depends on particular choice of coefficients
$C_n$, string masses $m_n$ and coupling constants $g_n$. To avoid a
divergence  in  $1/(n-1)$ when $n = 1$ one has to take that ${C_n\,
m_n^D}/{g_n^2}$ is proportional to $n -1$. Here we shall consider
some cases when coefficients $C_n$ are proportional to $n-1$, while
masses $m_n$ as well as coupling constants $g_n$ do not depend on
$n$, i.e. $ m_n =  m , \,\, g_n = g$. Since this is an attempt
towards effective Lagrangian  of an adelic string it seems natural
to take mass and coupling constant independent on particular $p$ or
$n$. To differ this new field from a particular $p$-adic one, we use
notation $\phi$ instead of $\varphi$.

\bigskip

{\centering\subsection{THREE TYPES OF LAGRANGIANS \label{tt}}}

 We are going to consider three cases, which depend on the choice of coefficients $C_n , \, n \geq 1$.

\bigskip

{\centering\subsubsection{Case $C_n = \frac{n - 1}{n^{2+h}}$
\label{sub}}}

Let us first consider  the case

\beq C_n = \frac{n-1}{n^{2+h}} \,, \label{2.5} \eeq where $h$ is a
real number. The corresponding Lagrangian is

\bea L_{h} = C_0 \, {\cal L}_0  +  \frac{m^D}{g^2} \Big[ -
\frac{1}{2}\, \phi \, \sum_{n= 1}^{+\infty} n^{-\frac{\Box}{2 m^2}
-h} \, \phi  + \sum_{n= 1}^{+\infty} \frac{n^{-h}}{n + 1} \,
\phi^{n+1} \Big] \label{2.6} \eea and it depends on parameter $h$.

According to the famous Euler product formula, one can write
$$ \sum_{n= 1}^{+\infty} n^{-\frac{\Box}{2\, m^2}  - h} = \prod_p \frac{1}{ 1 -
p^{-\frac{\Box}{2\, m^2}   - h}}\,. $$  Recall that standard
definition of the Riemann zeta function is

\beq  \zeta (s) = \sum_{n= 1}^{+\infty} \frac{1}{n^{s}} = \prod_p
\frac{1}{ 1 - p^{- s}}\,, \quad s = \sigma + i \tau \,, \quad \sigma
>1\,, \label{2.7} \eeq
which has analytic continuation to the entire
complex $s$ plane, excluding the point $s=1$, where it has a simple
pole with residue 1. Employing definition (\ref{2.7}), we can
rewrite (\ref{2.6}) in the form

 \beq L_{h} = C_0 \, {\cal L}_0  + \frac{m^D}{g^2} \Big[ \, -
\frac{1}{2}\,
 \phi \,  \zeta\Big({\frac{\Box}{2\, m^2}  +
h }\Big) \, \phi  +   \sum_{n= 1}^{+\infty} \frac{n^{ - h}}{n + 1}
\, \phi^{n+1} \Big]\,. \label{2.8} \eeq
Here
 $\zeta\Big({\frac{\Box}{2\, m^2} + h}\Big)$ acts as a
pseudodifferential operator  \beq \label{2.9}
\zeta\Big({\frac{\Box}{2\, m^2}  + h }\Big)\, \phi (x) =
\frac{1}{(2\pi)^D}\, \int e^{ ixk}\, \zeta\Big(-\frac{k^2}{2\, m^2}
 + h \Big)\, \tilde{\phi}(k)\,dk \,,
 \eeq
where $ \tilde{\phi}(k) =\int e^{(- i kx)} \,\phi (x)\, dx$ is the
Fourier transform of $\phi (x)$. Lagrangian $L_0 $, with the
restriction on momenta $-k^2 = k_0^2 -\overrightarrow{k}^2
> (2 - 2 h)\, m^2 $ and field $|\phi | < 1$, is analyzed in \cite{dragovich3}. In
\cite{dragovich5}, we considered Lagrangian (\ref{2.8}) with
analytic continuations of the zeta function and the power series
$\sum \frac{n^{-h}}{n + 1} \, \phi^{n+1}$, i.e.

\beq L_{h} = C_0 \, {\cal L}_0  + \frac{m^D}{g^2} \Big[ \,-
\frac{1}{2}\,
 \phi \,  \zeta\Big({\frac{\Box}{2 \, m^2}  +
h }\Big) \, \phi    + {\cal{AC}} \sum_{n= 1}^{+\infty} \frac{n^{-
h}}{n + 1} \, \phi^{n+1} \Big]\,, \label{2.10} \eeq where
$\mathcal{AC}$ denotes analytic continuation.

 Nonlocal dynamics of this field $\phi$ is
encoded in the pseudodifferential form of the Riemann zeta function.
When the d'Alembertian is in the argument of the Riemann zeta
function we  say that we have zeta nonlocality. Accordingly, this
$\phi$ is a zeta nonlocal scalar field.

\bigskip

{\centering\subsubsection{Case $C_n = \frac{n^2 - 1}{n^{2}}$
\label{sub}}}

In this case Lagrangian (\ref{2.4}) becomes

\bea L = C_0 \, {\cal L}_0  +   \frac{m^D}{g^2} \Big[ -
\frac{1}{2}\, \phi \, \sum_{n= 1}^{+\infty} \Big( n^{-\frac{\Box}{2
m^2} + 1} \, + \, n^{-\frac{\Box}{2 m^2}} \Big) \, \phi  + \sum_{n=
1}^{+\infty} \, \phi^{n+1} \Big] \label{2.17} \eea and it yields

 \beq L = C_0 \, {\cal L}_0  + \frac{m^D}{g^2} \Big[ \, - \frac{1}{2}\,
 \phi \,  \Big\{ \zeta\Big({\frac{\Box}{2\, m^2}  -
 1}\Big)\, + \, \zeta\Big({\frac{\Box}{2\, m^2} }\Big) \Big\} \, \phi \,  + \,   \frac{\phi^2}{1 - \phi} \,
 \Big]\,. \label{2.18} \eeq

Some classical field properties of this Lagrangian are analyzed and
presented in \cite{dragovich6}.

\bigskip

{\centering\subsubsection{Case $C_n = \mu (n)\, \frac{n - 1}{n^{2}}$
\label{sub}}}

Here $\mu (n)$ is the M\"obius function, which is defined for all
positive integers and has values $1, 0, -1$ depending on
factorization of $n$ into prime numbers $p$. It is defined as
follows:

\begin{equation}
\mu (n)= \left \{ \begin{array}{lll} 0 , \quad &  n = p^2 m \\
(-1)^k , \quad & n = p_1 p_2 \cdots p_k ,\,\,  p_i \neq p_j  \\
1 , \quad & n = 1, \,\,  (k=0)\, .
\end{array} \right.
\label{2.19}
\end{equation}

The corresponding Lagrangian is

\begin{equation}
L_\mu =  C_0 \, {\cal L}_0  +   \frac{m^D}{g^2} \Big[ -
\frac{1}{2}\, \phi \, \sum_{n= 1}^{+\infty}\, \frac{\mu (n)}{
n^{\frac{\Box}{2 m^2}}} \,\phi   + \sum_{n= 1}^{+\infty} \frac{\mu
(n)}{n + 1} \, \phi^{n+1} \Big] \label{2.20}
\end{equation}

Recall that the inverse Riemann zeta function can be defined by
\begin{equation}
\frac{1}{\zeta (s)} = \sum_{n =1}^{+\infty}\, \frac{\mu (n)}{n^s},
\quad s=\sigma + i \tau , \quad \sigma > 1. \label{2.21}
\end{equation}

Now (\ref{2.20}) can be rewritten as
\begin{equation}
L_\mu =  C_0 \, {\cal L}_0  +   \frac{m^D}{g^2} \Big[ -
\frac{1}{2}\, \phi \,  \frac{1}{ \zeta\Big({\frac{\Box}{2
m^2}}\Big)} \,\phi + \int_0^\phi {\cal M}(\phi) \, d\phi\Big] ,
\label{2.22}
\end{equation}
where ${\cal M}(\phi) = \sum_{n= 1}^{+\infty} {\mu (n)} \, \phi^{n}
= \phi - \phi^2 - \phi^3 - \phi^5 + \phi^6 - \phi^7 + \phi^{10} -
\phi^{11} - \dots $.

The corresponding potential, equation of motion and mass spectrum
formula, respectively, are:
\begin{equation}
V_\mu (\phi) = - L_\mu (\Box = 0) = \frac{m^D}{g^2} \Big[
\frac{C_0}{2}\, \phi^2 (1 - \ln \phi^2) - \phi^2 - \int_0^\phi {\cal
M}(\phi) \, d\phi \Big]\, , \label{2.23}
\end{equation}

\begin{equation}
\frac{1}{ \zeta\Big({\frac{\Box}{2 m^2}}\Big)} \,\phi - {\cal
M}(\phi) - C_0 \, \frac{\Box}{m^2}\, \phi - 2 C_0\, \phi \, \ln \phi
= 0\,, \label{2.24}
\end{equation}

\begin{equation}
\frac{1}{ \zeta\Big({\frac{M^2}{2 m^2}}\Big)} - C_0 \,
 \frac{M^2}{m^2} + 2 C_0 - 1 = 0 \,, \quad |\phi|\ll 1\,,
 \label{2.25}
\end{equation}
where usual relativistic kinematic relation $k^2 = - k_0^2
+\overrightarrow{k}^2 = - M^2$ is used.

Analysis of the above expressions will be presented elsewhere.

\bigskip

{\centering\subsubsection*{Acknowledgements}}

The work on this article was partially supported by the Ministry of
Science, Serbia, under contract No 144032D.

\bigskip

\begin{center}

\end{center}
\end{document}